\documentclass[conference]{IEEEtran}
\IEEEoverridecommandlockouts
\usepackage{cite}
\usepackage{booktabs}
\usepackage{amssymb,amsfonts}
\usepackage[fleqn]{amsmath}
\usepackage{graphicx}
\usepackage{mathdots}
\usepackage{comment}
\usepackage{xcolor}
\usepackage[classicReIm]{kpfonts}
\usepackage{mathptmx}
\usepackage{amssymb}
\usepackage[T1]{fontenc}
\usepackage{algorithm}
\usepackage{algorithmic}
\usepackage{amsmath}
\usepackage{graphicx}
\usepackage{longtable}
\usepackage{tabularray}
\usepackage{tabularx}
\definecolor{Gray}{gray}{0.75}
\usepackage{multirow}
\usepackage{xcolor,colortbl}
\definecolor{green}{rgb}{0,1,0}

\def\BibTeX{{\rm B\kern-.05em{\sc i\kern-.025em b}\kern-.08em
    T\kern-.1667em\lower.7ex\hbox{E}\kern-.125emX}}
\begin{document}

%\title{Asymmetric Byzantine Quorum Approach to Resolve Trust Issues in Blockchain Decentralized Oracle}
% \title{Addressing Trust Challenges in Decentralized Blockchain Oracles using Asymmetric Byzantine Quorum}
\title{Addressing Trust Challenges in Blockchain Oracles using Asymmetric Byzantine Quorums}

\begin{comment}
\author{\IEEEauthorblockN{Fahad Rahman}
\IEEEauthorblockA{\textit{}
\textit{Université Paris Cité}\\
fahad.rahman@etu.u-paris.fr}
\and

\IEEEauthorblockN{Chafiq Titouna}
\IEEEauthorblockA{\textit{}
\textit{LIGM, ESIEE, University Gustave Eiffel}\\
chafiq.titouna@esiee.fr}
\and

\IEEEauthorblockN{Farid Naït-Abdesselam}
\IEEEauthorblockA{\textit{}
\textit{University of Missouri Kansas City}\\
naf@umkc.edu}
}
\end{comment}

\author{
    \IEEEauthorblockN{
        Fahad Rahman\IEEEauthorrefmark{1},
        Chafiq Titouna\IEEEauthorrefmark{2} and
        Farid Naït-Abdesselam\IEEEauthorrefmark{3}
        } 
\IEEEauthorblockA{\IEEEauthorrefmark{1}Université Paris Cité, France\\}
\IEEEauthorblockA{\IEEEauthorrefmark{2}LIGM, ESIEE Paris, University Gustave Eiffel, France\\}
\IEEEauthorblockA{\IEEEauthorrefmark{3}University of Missouri Kansas City, USA}\\
    }
    
\maketitle
\noindent
\begin{abstract}
Distributed computing in Blockchain Technology (BCT) hinges on a trust assumption among independent nodes. Without a third-party interface or what's known as a `Blockchain Oracle', it can't interact with the external world. This Oracle plays a crucial role by feeding extrinsic data into the Blockchain, ensuring that Smart Contracts operate accurately in real time. The `Oracle problem' arises from the inherent difficulty in verifying the truthfulness of the data sourced by these Oracles. The genuineness of a Blockchain Oracle is paramount, as it directly influences the Blockchain's reliability, credibility, and scalability. To tackle these challenges, a strategy rooted in Byzantine fault-tolerance $\phi$ is introduced. Furthermore, an autonomous system for sustainability and audibility, built on heuristic detection, is put forth. The effectiveness and precision of the proposed strategy outperformed existing methods using two real-world datasets, aimed to meet the authenticity standards for Blockchain Oracles.

\end{abstract}
\section{Introduction}
\par
The BCT operates as a unified system of distributed ledgers underpinned by consensus mechanisms. Without an intermediary or interface, it remains disconnected from the external environment. Any data recorded on a single node is automatically mirrored across the entire Blockchain network. Digital agreements stored within Blockchain nodes are termed `Smart Contracts'~\cite{hewa2021survey} These are self-executing code segments that activate based on specific inputs/ outputs, running autonomously within the Blockchain\cite{permenev2020verx}. The external data channeled into Smart Contracts is termed as `Blockchain Oracle' \cite{wang2019blockchain}~\cite{gupta2019layer}. 

The name `Oracle' isn't tied to a particular device (like IoT) or software. Drawing from Greek mythology, an `Oracle' was an individual or entity believed to have a direct line to the divine, offering predictions of the future. Historically, Oracles served as a source of knowledge beyond human understanding, guiding those who lacked the necessary information to decide~\cite{mammadzada2020blockchain}~\cite{leite2021assuring}. As depicted in Fig. 1, a Smart Contract relies on authentic extrinsic data to operate, 
\begin{figure}[ht]
\centerline{\includegraphics[width=0.4\textwidth]{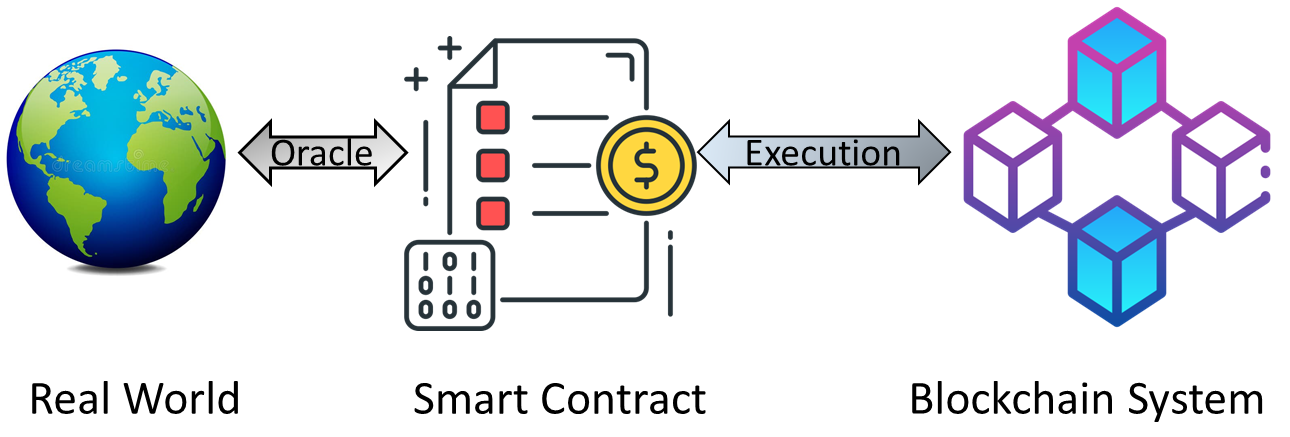}}
\caption{Data transition, real-world to Blockchain}
\label{fig1}
\end{figure}
The significance of studying Blockchain Oracles stems from their pivotal role in linking Blockchain mechanisms, especially Smart Contracts, with real-world data. Various Blockchain-driven systems, including prediction markets for forecasting, currency exchange platforms, sports betting, and weather reporting applications~\cite{mcghin2019blockchain}~\cite{sunny2020supply}~\cite{mohammedali2023management}, some salient applications of Blockchain Oracle are mentioned in Table I.

At the heart of the Blockchain Oracle dilemma is the need to guarantee that data sourced from the external environment and fed into a Smart Contract through a Blockchain Oracle is both trustworthy and unquestionably reliable for all Smart Contract stakeholders\cite{yang2022method}. If a Blockchain Oracle provides erroneous or misleading information, it risks undermining the entire Blockchain system's trustworthiness. Within a distributed network, even if most nodes operate with integrity, a minority presenting compromised~\cite{garagad2022dynamic}, inaccurate, or biased values can introduce inconsistencies in Smart Contract outcomes. Pinpointing a rogue node based on its output is also part of the Oracle challenge~\cite{jebari2022analysis}. Additionally, scalability, measured by transactions per second (TPS), remains a prominent concern in both public\cite{lai2018blockchain} and private Blockchains\cite{nguyen2019optchain}. The operational efficiency of a Blockchain Oracle significantly influences overall Blockchain scalability~\cite{gupta2022secure}.

\begin{table*}[t]
\caption{Use of decentralized oracle in real-world applications}
\begin{tabular}{|p{5cm}|p{6cm}|p{6cm}|}

\hline
\centering Application Area & \centering Description & Use of Oracles \\ 
\hline
Supply Chain Management\cite{baktayan2022intelligent}~\cite{tahmasbi2023identifying} & Sensors monitor conditions like temperature, humidity, and location during product shipment. & Ensures data integrity and transparency on product history, reducing disputes among parties  \\ 
Smart Agriculture\cite{belen2023systematic} & Sensors track soil moisture, weather, and crop health. & Automates farming activities and insurance payouts, based on reliable, real-time data  \\ 
Energy Grid Management\cite{han2022gap}  & Sensors record energy production and consumption in decentralized grids. & Balances supply, demand, and pricing in real-time, enabling transparent energy trading  \\ 
Environmental Monitoring\cite{shodiq2021secure} & Sensors worldwide track pollution levels, temperature, and deforestation. & Provides tamper-proof environmental data for reporting and carbon credit trading  \\ 
Healthcare Monitoring\cite{attaran2022blockchain} &Wearable devices monitor patient health metrics. & Updates Blockchain records securely, ensuring accurate, immutable medical data for remote healthcare  \\ 
Automated Insurance\cite{amponsah2022improving}  & Sensors detect conditions meeting insurance claim criteria, like car accidents or home damage. & Triggers automatic, tamper-proof claims processing and payouts  \\ 
Smart Property Management~\cite{mededjel2022blockchain} & Building sensors monitor occupancy, temperature, and security. & Automates building management, energy efficiency, and lease agreements  \\ 
Quality Assurance in Manufacturing~\cite{alessi2019decentralized} & Production line sensors identify defects or equipment issues. & Triggers quality control and maintenance, ensuring product standards  \\ 
Traffic and Urban Planning~\cite{cengiz2021role} & City-wide sensors gather traffic, parking, and public transport data. & Informs urban planning and automates traffic management or toll payments \\ 
Seismic Activity Reporting~\cite{r2020evaluation} & Seismic sensors detect geological events. & Enables rapid, reliable data recording for early warnings and emergency responses  \\ 
\hline
\end{tabular}
\label{Table-3: literature review}
\end{table*}

A scalable and precise Blockchain Oracle is crucial to avert lags, inaccuracies, and possible security vulnerabilities when processing Smart Contracts dependent on extrinsic information. This study presents the Asymmetric Byzantine Quorums (ABQ) method, which fully supports Byzantine fault-tolerance, aiming to promptly ascertain trustworthy and accurate Blockchain Oracle values. Additionally, we've integrated a heuristic-driven detection system to pinpoint malicious entities and ensure traceability. Our proposed methodology is apt for both Public and Private Blockchains. The model offers extensive potential for data aggregation in Oracle, suitable for a multitude of real-world decentralized Blockchain Oracle applications. 

The structure of this paper is outlined as follows: Section II delves into relevant literature and related work. The system model is concisely introduced in Section III, followed by a detailed discussion of the methodology in Section IV. Section V assesses simulations using real-world datasets, while Conclusion and Future Work are discussed in Section VI. 

\section{Related Work}
The challenge of the Blockchain Oracle has been tackled in various research papers. This section highlights some notable contributions: Ellis et al., in their work \cite{ellis2017chainlink}, introduced a weightage-based Oracle approach. Each IoT/ input stream was assigned a specific weightage for Oracle input. Problems arise if the $IoT$ device with the highest repute provides notably different data to Oracle, jeopardizing the entire Blockchain system's integrity. In another study, Adler et al. \cite{adler2018astraea} approached the issue of Oracle through 'Game-theory'. They presented a scheme of bi-layered voting: the first layer is comprised of voters, while the second one consists of certifiers. Voters received lesser rewards compared to certifiers. If a certifier detects discrepancies in the voters' results, they would earn a substantial reward. While this system often produced accurate outputs, vulnerabilities arose if a certifier was compromised or if a certifier and voter colluded to provide false data, compromising the system's integrity.

Tian et al. \cite{tian2019evaluating} noted the challenge of anticipating attacker strategies due to the attackers' varied nature. Attackers lack knowledge of the comprehensive reputation management system and the readings from competing IoT devices. The authors used both entity and data-centric schemes to devise a foundational trust management computation model. In this framework, each vehicle and traffic event notification possessed distinct reputation values. Those with zero reputation were deemed unreliable and excluded from the system and broadcast lists. This reputation mechanism was anchored in Game theory principles, with nodes of higher reputation posing greater risks. Lastly, Heiss et al. \cite{heiss2019oracles} put forth a voting-centric system. Here, the value of every $IoT$ device received votes, and the value with the highest vote is deemed accurate. A designated time window existed for vote submission, followed by automated vote tallying. Voters were rewarded for accurate values. The system also featured weighted voting, wherein rewards and penalties were determined based on Game theory principles. However, the system faced vulnerabilities if highly-voted nodes were hacked or malfunctioned.

\begin{table*}[t]
\caption{Existing related work}
\begin{tabular}{|p{2cm}|p{1.5cm}|p{1.5cm}|p{4.5cm}|p{2.5cm}|p{2.5cm}|}
\hline
Authors & Scalability & Oracle Type & Methodology &  Strengths & Weaknesses\\ 
\hline
S. Ellis et al.~\cite{ellis2017chainlink} & Yes & Centralized & Weighted scoring approach & Fast processing & Low reliability \\ 
Adler et al.~\cite{adler2018astraea} & No & Centralized & Cross-layer approach & Game theory & Complex   \\ 
Tian et al.~\cite{tian2019evaluating} & No & Centralized & Reputation-based system & Fast processing & Low reliability   \\ 
% Gautam et al.~\cite{srivastava2019light}            & Yes & Yes  & No & No & Medium  \\ 
Heiss et al.~\cite{heiss2019oracles} & No & Centralized & Voting-based system & Game theory & Low reliability   \\ 
Berger et al.~\cite{berger2020aware} & Yes & Centralized & Geographical scalability & Fast processing & Low reliability   \\ 
Tseng et al.~\cite{tseng2020exact} & No & Centralized & Synchronous and Asynchronous model & Fast processing & Entity integrity   \\ 
Wei et al.~\cite{wei2020pspl} & No & Centralized & Neighbour
discovery algorithm &  Low energy & Data overlapping   \\ 
\hline
\end{tabular}
\label{Table-3: literature review}
\end{table*}

Berger et al., in their study \cite{berger2020aware}, introduced the 'Adaptive Wide-Area REplication' (AWARE) approach. This approach aims to enhance the geographical scalability of consensus among nodes dispersed across vast physical distances. The approach integrates a voting and weightage-based model to form distinct 'Asymmetric Quorums'. The system's integrity and dependability are influenced by the weightage and quantity of nodes, guiding the creation of quorums. In another work by Tseng et al. \cite{tseng2020exact}, two communication models were put forth: Synchronous and Asynchronous. In the synchronous model, communication unfolds in cycles, with each node following a specific sequence. On the other hand, the asynchronous model lacks a predefined communication sequence, allowing nodes to operate randomly and messages to be delayed unpredictably. Nodes in this model communicate with their immediate neighbors via established, dedicated ports. A vulnerability within this system is that if a transmitting device is duplicated and then dispatches malicious data under the original identity, the entire system's security is jeopardized.

Wei et al., in their paper \cite{wei2020pspl}, proposed a dual-direction neighbor discovery mechanism for IoT devices, although its foundation is based on single-direction discovery. The duration of active slots for devices has been minimized, resulting in reduced energy consumption when identifying new IoT neighbors. For neighbor discovery, they employed an asymmetric neighboring model using `Pure-Transmitting' (PS) and `Pure-Listening' (PL) intervals. The overlapping approach discussed in their paper offers a superior solution to the challenges they addressed.

A review of the aforementioned studies and Table II reveals a gap: there is no method that effectively filters out malicious or compromised data. Furthermore, a single erroneous data entry can have catastrophic repercussions for the entire system. Similarly, techniques based on voting and weighting can occasionally exacerbate issues.

\section{System Model}
We introduced the ABQ approach to ensure accurate, reliable, and scalable Oracle for the Smart Contract/ Blockchain. Consider a large-scale farm, equipped with various IoT devices to monitor soil moisture, temperature, and humidity levels for optimized irrigation. The objective is to automate irrigation based on real-time data from IoT devices using a Blockchain-powered Smart Contract. The Blockchain ensures data integrity, and the Smart Contract ensures that water is released only when required. In this scenario ‘Actors’ are IoT devices (soil moisture sensors, temperature sensors, humidity sensors), Blockchain Oracle, Blockchain network with Smart Contracts, Farm owner or manager, and Watering system.

Initially, the farm owner deploys a Smart Contract on the Blockchain. This contract is designed to receive data from Oracle and execute the irrigation process based on predefined conditions. IoT devices are set up across the farm and connected to an IoT platform that collects and sends data to the Oracle. Every hour, IoT devices collect data about soil moisture, temperature, and humidity. This data is sent to the IoT platform. The Oracle retrieves data from the IoT platform. To ensure data reliability, it can pull data from multiple sources or verify it against multiple similar IoT devices. The Oracle then sends this verified data to the Blockchain. Once the Smart Contract on the Blockchain receives the data from Oracle, it checks the conditions. For instance, if soil moisture is below  50\% and the temperature is above 30°C, then activate the watering system, if the humidity is above 80\%, then delay watering regardless of other conditions. Based on the conditions set in the Smart Contract and the data received from the Oracle, the Smart Contract sends a command to the watering system to start or stop the irrigation process. After irrigation, IoT devices will continue to monitor the conditions. If the soil moisture reaches an acceptable level, the Smart Contract may send a command to stop the watering. This creates a feedback loop ensuring optimal watering based on real-time conditions. 

All transactions (data inputs and irrigation commands) are recorded on the Blockchain, ensuring transparency. The farm owner or manager can review the Blockchain records to verify that the irrigation was done based on actual field conditions. This helps in building trust in the automated system. By using Blockchain, the farm owner has a transparent and tamper-proof record of all actions taken by the system. Water resources are used more efficiently, leading to cost savings and sustainable farming. In this scenario the main challenge is the reliability of Blockchain Oracle, the input IoT devices can malfunction/be compromised or hacked. It's crucial to have reliable and scalable Oracle to ensure that Oracle sends accurate data to Smart Contract. In the case of complex Oracle calculations, there might also be delays in data transmission and processing from the Oracle to the Blockchain. This is also accounted for in the proposed Blockchain Oracle model.

Our suggested Oracle system offers input from the external environment (outside Blockchain) by gathering temperature and other readings from IoT devices. These readings trigger the Smart Contract's execution. Our algorithm filters out any erroneous or compromised readings from the IoT devices, gathering only authentic readings into a data array. Correct readings from various IoT devices, captured at a particular moment, form the ABQ. The quorum's average becomes the definitive value for the Oracle~\cite{10271588}, as depicted in Fig. 2. 

\begin{figure}[b]
\centerline{\includegraphics[height=3.75cm,width=0.5\textwidth]{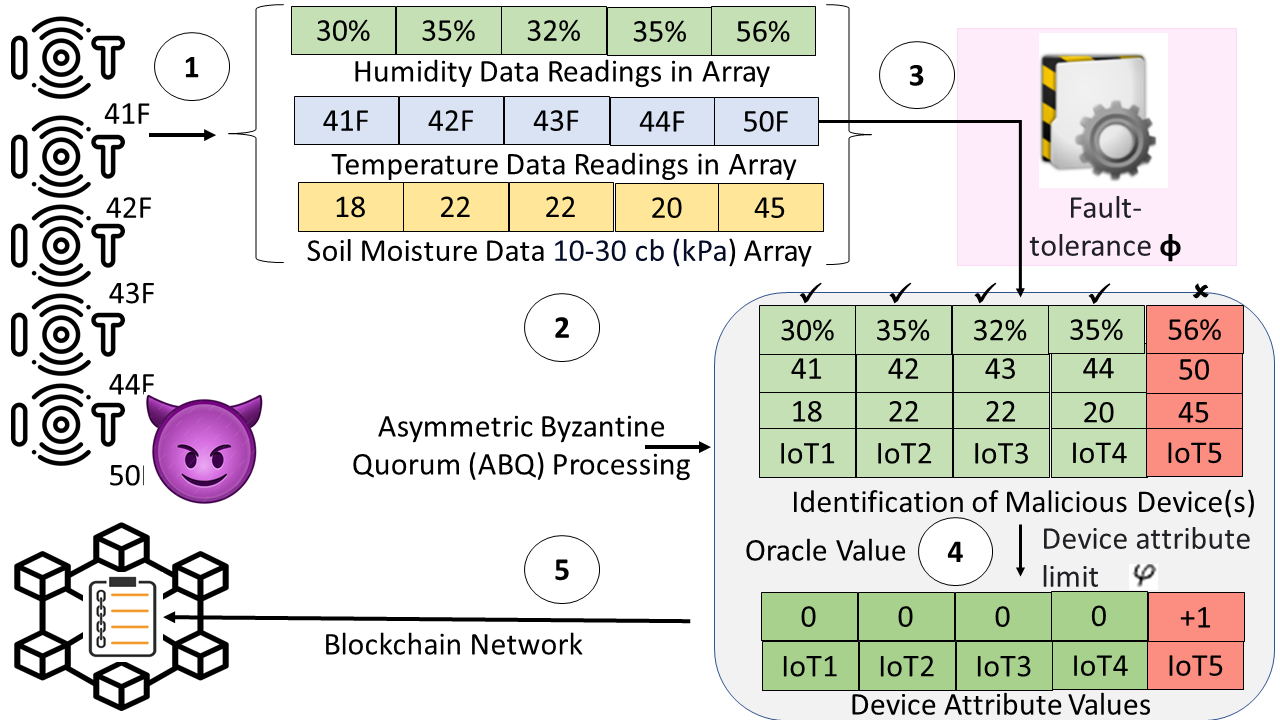}}
\caption{Data transmission and processing: Field to Blockchain through Oracle}
\label{fig2}
\end{figure}

The aim of this study is to address external influences, such as hacking attempts or errors in transmission/ data processing, that might compromise Oracle data. With the described technique, data from $IoT devices$ undergoes processing across distinct units. Each of these units aggregates readings within a specified time frame to data quorum. Sudden deviations/ irregularities in the readings are instantly detected and removed within the quorums. The proposed Oracle plays an intermediary role between sensors/ $IoT devices$ and Smart Contracts. 
% This layer is situated in an impartial location, selected through joint agreement by all stakeholders participating in the Smart Contract.

\section{Methodology}

In this study, a method that leverages a Byzantine Fault Tolerant (BFT) strategy using 'Asymmetric trust'~\cite{lim2023trust} to filter out deceptive readings from IoT devices, ensuring a consistent and authentic value of Blockchain Oracle is introduced. Data acquired from IoT devices is organized into an array. This data undergoes evaluation based on a pre-set fault-tolerance $\phi$ threshold to establish the proposed $ABQ$. The collected data, accumulated at regular time intervals from various $IoT-devices$ within a designated time frame, constitutes the quorum(s). A quorum is derived from the IoT data that satisfies the $\phi$ conditions after the spontaneous removal of malicious entries. For instance, if the $\phi$ threshold is set at $2$ and the discrepancy between the chosen median and the data values surpasses this threshold ($\phi$ $>$ $2$) that particular value will be automatically discarded. Any data value falling outside the $\phi$ range is likely to be erroneous or malicious. This study is segmented into 2 primary sections:
\begin{enumerate}
\item [A.] Establishment of $ABQ$ and Determining Oracle Values
\item [B.] Detection of faulty/ compromised device(s)
\end{enumerate}
\subsection{Establishment of ABQs and Determining Oracle Values}

The basic premise of BCT is predicated on the idea that at least two-thirds $(2/3)$ of its participants are acting truthfully. In this study, we have defined an ABQ as an asymmetric aggregation of data units, consisting of a set that exceeds half the sum of the total number of processes ($N_p$) and the number of faulty processes ($f_p$), mathematically expressed as {({$N_p$+$f_p$})/{2}}. The establishment of a BQ adheres to three core proposed assumptions, with the first being fundamental and the subsequent two being consequential derivations~\cite{alpos2021trust}.

\subsubsection{BQ Primary ($1^{st}$) Property}
At least one ($BQ$) exists containing solely accurate readings, as illustrated in Fig. 3 a. A quorum can be characterized by its correct processes. Values within a particular quorum that follow a specific order are deemed correct based on primary ($1^{st}$) property,

\subsubsection{BQ's Secondary ($2^{nd}$) Property}
Any pair of BQs should have an intersection that encompasses at least one accurate value. When two divergent clusters overlap, the common value they share, as depicted in Fig. 3 b, stands as the authentic and trustworthy value affirmed by both sets. It's important to highlight that as the frequency of quorum intersections on a singular value increases (within the same time frame), the resultant values exhibit enhanced accuracy. For any two quorums $Q_1$ and $Q_2$ in the quorum system, there must be at least one node that belongs to both $Q_1$ and $Q_2$.
Equation 1, for all $Q_1$ and $Q_2$ in the quorum system:
\begin{align}
Q_1 \cap Q_2 & \neq  \varnothing 
\end{align}
This property is essential for reaching consensus and preventing conflicts in the system. The quorum system must ensure that any two quorums that intersect (satisfying the Intersection property) must also have at least one node in common that is non-faulty. Equation 2, for all $Q_1$ and $Q_2$ in the quorum system where $Q_1$ 
 $\cap$ $Q_2$ $\neq$ $\varnothing$:
\begin{align}
|Q_1 \cap Q_2| > F_n  
\end{align}
In Equation 2, the $F_n$ represents the set of Byzantine faulty nodes. This property ensures that even if two quorums intersect, they must still have at least one non-faulty node in common, preventing Byzantine nodes from causing conflicts.

% \begin{figure}[t]
% \centerline{\includegraphics[height=4cm,width=0.4\textwidth]{./Images/3-properties-clr.png}}
% \caption{Three Characteristics of BQ}
% \label{fig1-3}
% \end{figure}
\begin{figure}[t]
\centerline{\includegraphics[height=5.2cm,width=0.5\textwidth]{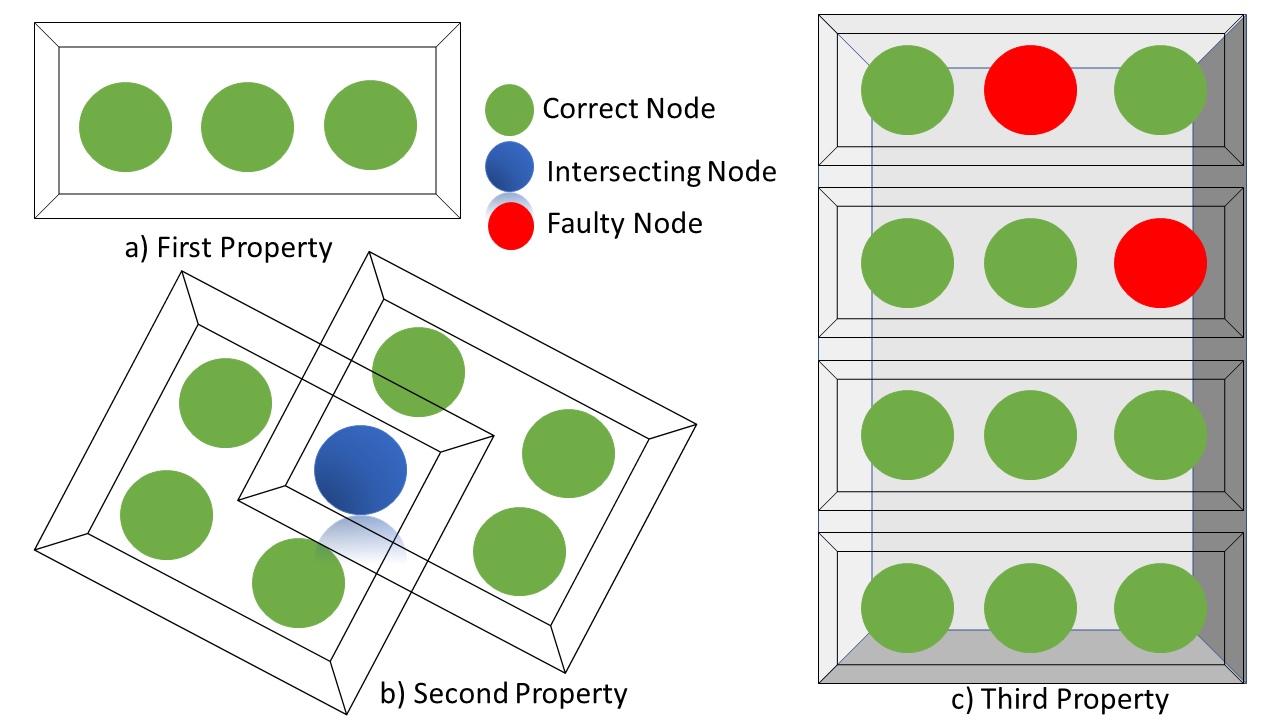}}
\caption{Three Characteristics/ properties of BQ}
\label{fig1-3}
\end{figure}

\subsubsection{BQ's Tertiary ($3^{rd}$) Property}
Within a BQ, the predominant values must be accurate, as illustrated in Fig. 3 c. The distributed system inherently operates on the assumption that most of the nodes, data sources, sensors, or input streams function correctly and contribute to the stability of the system. If the majority were to act maliciously, the viability of the distributed system would be jeopardized. The quorum system must be able to tolerate a certain number of faulty nodes. Specifically, the number of nodes in the quorum system should be greater than three times the maximum number of Byzantine faulty nodes $3(Fm)$, where $F_m$ is the maximum number of faults the system is designed to handle. Equation 3, for all $Q$ in the quorum system and for any $F_m$ $\leq$ $n_n$/3 (where $n_n$ is the total number of nodes in the system):
\begin{align}
|Q| > 3F_m 
\end{align}
This property ensures that there are enough non-faulty nodes in each quorum to overcome Byzantine faults and maintain the integrity of the system.

In addition to $ABQ$, our method incorporates the longest-chain rule\cite{shi2019analysis} to derive precise Oracle values, which is a prevalent approach in BCT. This rule mandates that nodes recognize and accept the lengthiest chain among all candidates. If nodes opt for honest behavior, they should either support chains that match the length of previously broadcasted honest chains or exceed them\cite{blum2020combinatorics}. In our context, we perceive a quorum as a data chain. If a quorum gets fragmented into multiple segments, the lengthiest data segment is identified as the `longest-chain' which is deemed accurate, as depicted in Fig. 4. This approach yields a genuine and reliable data reading since it also aligns with the {($N_p$+$f_p$)/2} BQ definition. For instance, if $Q_1$ represents a quorum of number of nodes $(n_n)$, $Q_2$ is a quorum of $n_n+1$ and $Q_3$ is a quorum of $n_n+2$ reading, then given that $Q_3 > Q_2 > Q_1$, the $Q_3$ quorum is preferred over other quorums.

\begin{algorithm}[b]
\caption{Forming ABQs and Determining Oracle Value}
\begin{algorithmic} 
\REQUIRE  
{Input $\leftarrow$} IoT Device Stream $(IoT_1, IoT_2, \ldots IoT_n)$ \\ {Input $\leftarrow$} Data Generated by Devices $(dt_1$ $\rightarrow$ $dt_n)$ \\ {Input $\leftarrow$} Error Margin (Fault-tolerance): $\phi \in \mathbb{Z}$ \\ 
\STATE Duration Period $Tm_i$ Save data to an array ($Ar_i$)
\STATE For each data fetching $Tm_i$ 
\STATE Record data in an array ($Ar_i$)
\STATE Filtered Readings data in an array ($Rd$)
\IF {($Ar_i$) $\neq$ Sorted($Ar_i$)} 
    \STATE Sort $(Ar_i)$ in increasing sequence
\ENDIF
\STATE \textbf{Med = Med($Ar_i$)}
\STATE Rd=[(Med - $d_1$), (Med - $d_2$),...(Med - $d_n$)]
\STATE \textbf{Abslt (Rd)}
\IF {(|Rd| fetched readings ) $\leq$ $\phi$} 
    \STATE Approve the |Rd| readings [($d_1), (d_2) \ldots (d_n$)]  
    \STATE {Quorum ($Q_m$)  $\leftarrow$ Approve |Rd| readings [($dt_1), (dt_2) \ldots (dt_n$)]}
    \STATE \textbf{Oracle = Mean-Readings ($Q_m$)}
\ENDIF 
\STATE \textbf{end}
\end{algorithmic}
\end{algorithm}

% \begin{figure}[t]
% \centerline{\includegraphics[height=4cm,width=0.4\textwidth]{./Images/list-clr.png}}
% \caption{Rule of the Longest Chain for Quorum Formation}
% \label{fig4.1}
% \end{figure}
\begin{figure}[t]
\centerline{\includegraphics[height=5cm,width=0.5\textwidth]{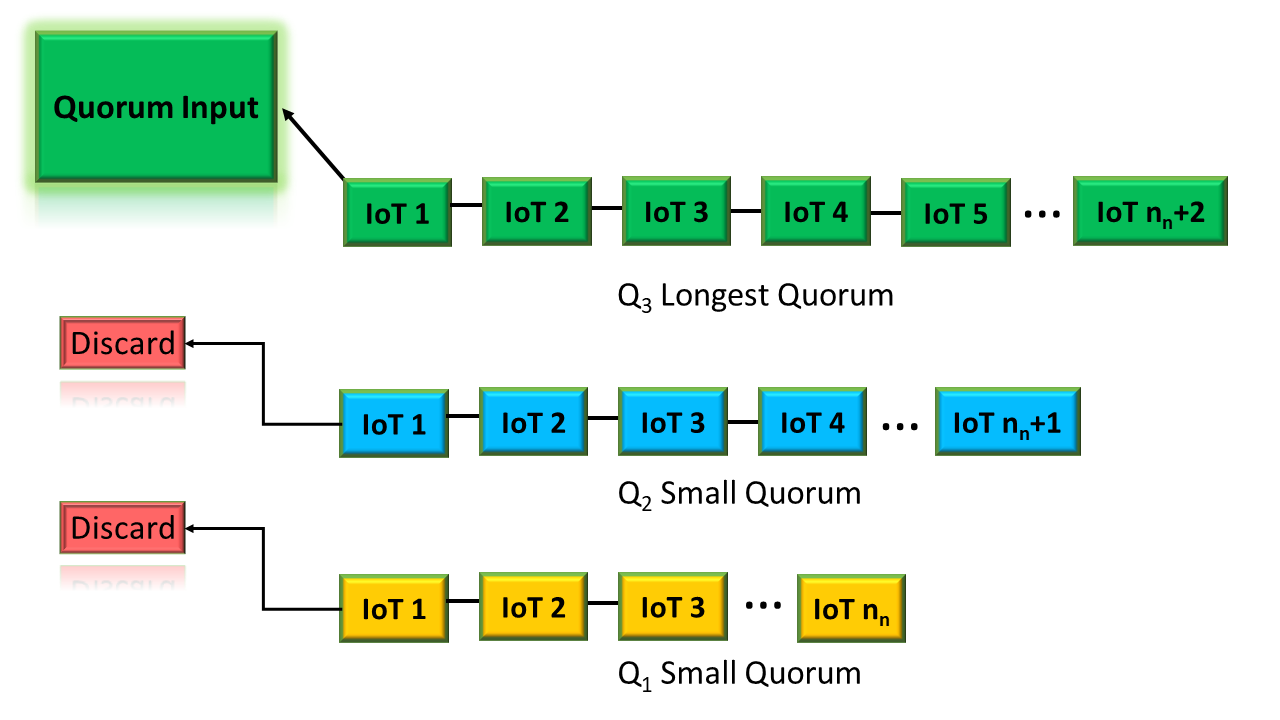}}
\caption{Rule of the Longest Chain for Quorum Formation}
\label{fig4.1}
\end{figure}

In this work, to implement the initial segment the input streams mentioned below for the formation of ABQs and determination of the Oracle are required:
i) Array containing data from $n_t$ number of IoT devices
ii) A predetermined threshold $\phi$. A lower threshold $\phi$ yields more accurate results. This threshold $\phi$ can be computed using the formula  ``$\phi$ = $m$ / $n_t$", where $m$ represents the permissible number of subsystem failures and $n_t$ is the total count of subsystems/ $IoT$ devices. Algorithm 1 delineates the process for forming ABQs and determining the Oracle value. The foundational requirements of our proposed system encompass a total of $n$ IoT devices and a predetermined $\phi$. The system's operation unfolds as follows: i) IoT device readings are captured and stored in an array accompanied by a timestamp. ii) If not already sorted, the data is arranged in ascending order. iii) The median of the data array is calculated. The difference between the median and each individual data value is computed. Given that the difference between the median and data values can result in negative numbers, the absolute $Abs$ value of the results is taken. The $\phi$ value is then applied to the $Abs$ outcomes. Consequently, values less than or equal to $\phi$ are deemed valid and are stored in a separate array. A quorum is then constituted using these accepted values. The final value of the Blockchain Oracle is determined by computing the mean of the quorum values. A state transition diagram illustrating this process is provided in Fig. 5.

\begin{figure*}[t]
\centerline{\includegraphics[height=7cm,width=0.7\textwidth]{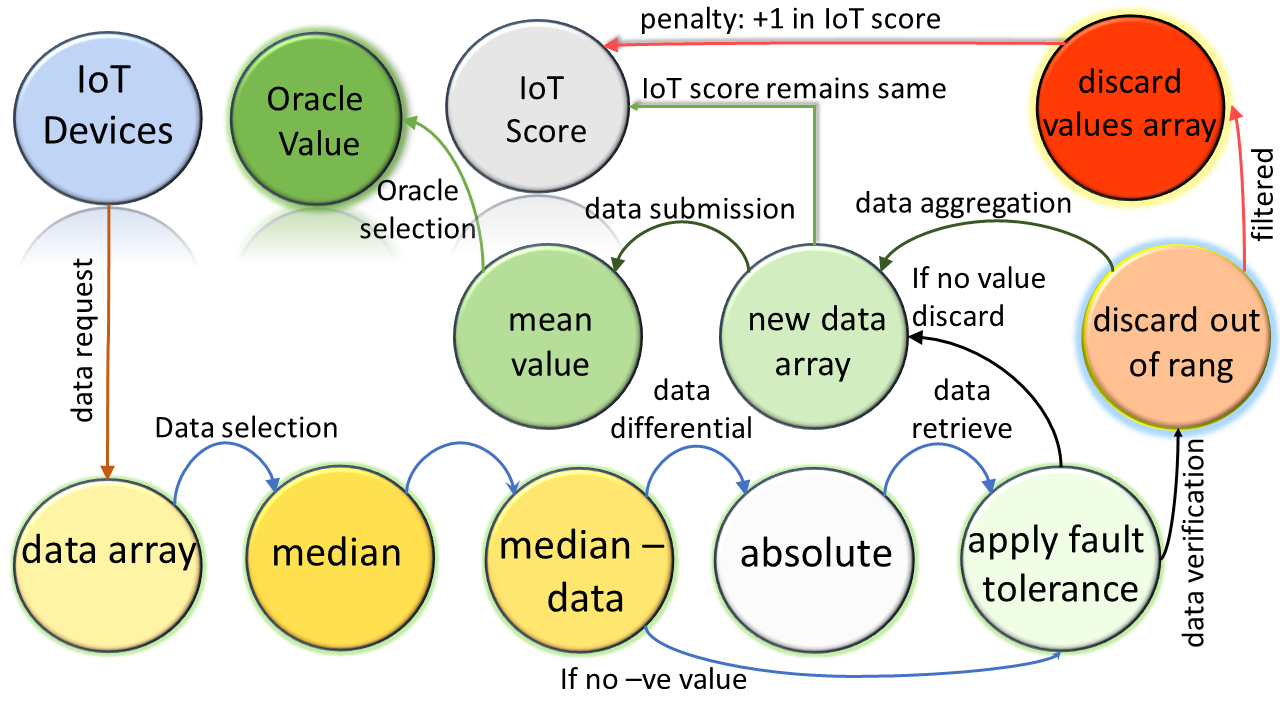}}
\caption{State transition diagram of proposed model}
\label{fig1-3}
\end{figure*}

% \begin{center}
% \begin{tabular}{ |c|c|c|c| } 
% \hline
% col1 & col2 & col3 \\
% \hline
% \multirow{3}{4em}{Multiple row} & cell2 & cell3 \\ 
% \hline
% & cell5 & cell6 \\ 
% \hline
% & cell8 & cell9 \\ 
% \hline
% \end{tabular}
% \end{center}
Let $IoT_1, IoT_2, \ldots IoT_n$ be the pointers of $IoT$ devices having by default score $\Psi$ = $0$ that are remotely connected to the system and transmitting data $dt_1, dt_2, \ldots dt_n$, establish a predefined $\phi$ value, denoted as {$\phi \in \mathbb{Z} $}. Capture the data in an array $A=[dt_1, dt_2, \ldots dt_n$]. Next, sort array $A$ in ascending order and determine its median $Med$. Subsequently, compute $D$ which represents the absolute difference between each transmitted data value from $Med$, values that are less than or equal to $\phi$ are deemed valid. The $Mean$ of these valid values is taken to form the Quorum $Q_m$.
\begin{table}[ht]
\caption{Oracle calculation steps in the proposed model}
\noindent \begin{tabularx}{0.485\textwidth}{ 
 | >{\centering\arraybackslash}X 
 | >{\centering\arraybackslash}X 
 | >{\centering\arraybackslash}X 
 | >{\centering\arraybackslash}X 
 | >{\centering\arraybackslash}X 
 | >{\centering\arraybackslash}X 
 | >{\centering\arraybackslash}X 
 | >{\centering\arraybackslash}X 
 | >{\centering\arraybackslash}X 
 | >{\centering\arraybackslash}X 
 | >{\centering\arraybackslash}X | }
 \hline
 & $  \cellcolor{gray} d_1$ &  \cellcolor{gray} $d_2$ &  \cellcolor{gray} $d_3$ &  \cellcolor{gray} $d_4$ &  \cellcolor{gray} $d_5$ &  \cellcolor{gray} $d_6$ &  \cellcolor{gray} $d_7$ &  \cellcolor{gray} $d_8$ &  \cellcolor{gray} $d_9$ & $Ans$\\
 \hline
i & \cellcolor{orange} 10 & \cellcolor{orange} 10 & \cellcolor{orange} 7 & \cellcolor{orange} 10 & \cellcolor{orange} 10 & \cellcolor{orange} 10 & \cellcolor{orange} 11 & \cellcolor{orange} 13 & \cellcolor{orange} 19 &\\
 \hline
ii & 10 & 10 & 7 & 10 & \cellcolor{yellow} 10 & 10 & 11 & 13 & 19 & \\
 \hline
iii & 0 & 0 & 3 & 0 & 0 & 0 & -1 & -3 & -9 &  \\
\hline
iv & 0 & 0 & 3 & 0 & 0 & 0 & 1 & 3 & 9 & \\
\hline
v & 0 & 0 & \cellcolor{red} 3 & 0 & 0 & 0 & 1 & \cellcolor{red} 3 & \cellcolor{red} 9 & \\
\hline
vi & 10 & 10 & & 10 & 10 & 10 & 11 &  &  & \\
\hline
vii &  &  &  &  &  &  &  &  &  & \cellcolor{green} 10\\
\hline
\end{tabularx}  \\
\label{Table-2: reliability}
\end{table}

As mentioned in Table III, a total of seven steps are required ($i - vii$) to find $Oracle$ value by using the proposed approach, (where $\phi$ = $2$. In step $i$ we get the $data$ from different devices (IoTs), find $median$ in step $ii$, then find $difference$ between $median$ and $data\ values$ in step $iii$, apply Absolute in step $iv$, next apply $\phi$ value and discard out of ranged values in step $v$, put actual data values in remaining slots in step $vi$, subsequently take a $mean$ of values in step $vii$, it will be sent as $Oracle$ value. 

\begin{algorithm}[b]
\caption{Detection of Malicious Device(s)}
\begin{algorithmic} 
\REQUIRE 
{Input $\leftarrow$} $IoT$ devices-score $(D_s)$: $\Psi \in  \mathbb{Z}$, \\ Max. score limit $\rightarrow$ $\varphi$ e.g. 5 \\ Initialise $\rightarrow$ $\Psi \rightarrow 0$  \\ 
% Initialise $\rightarrow$ $\varphi$ score max-limit = 5 \\
\STATE $Dsct \gets$ discarded-value(s) 
\STATE $IoT \gets Dsct$
\STATE $IoT$ $\Psi ++$ for each $Dsct$
\STATE \textbf{if} $IoT$ $\Psi ++$ = $\varphi$ 
\STATE \hspace*{0.5cm} Suspend $(dt_1, dt_2\ldots dt_n) \gets (IoT_1, IoT_2 \ldots IoT_n$) 
\STATE Trigger alert 
\STATE \textbf{end}
\end{algorithmic}
\end{algorithm}

% \begin{figure}[t]
% \centerline{\includegraphics[height=5cm,width=0.5\textwidth]{./Images/mean1.png}}
% \caption{Formation of ABQs and finding Oracle Value}
% \label{fig7.1}
% \end{figure}  mean1 yesterdays's work

\begin{comment}
We compared three different detection techniques i.e. Signature-based detection\cite{masdari2020survey}, Behavioral-based detection\cite{zhao2021detecting}, and Heuristic-based detection to get the most suitable detection technique, as shown in Table I. 

\begin{table}[H]
\caption{Comparison of different detection techniques}
\begin{center}
\begin{tabular}{m{1.2in}>
{\centering\arraybackslash}m{0.5in}>{\centering\aybackslash}m{0.5in}>{\centering\arraybackslash}m{0.5in}>
{\centering\arraybackslash}m{0.5in}}
\hline  
Characteristics & Signature Based Detection & Behavioral Based Detection & Heuristic Based Detection \\  
\hline 
Data Checking &$\times$ & \checkmark & \checkmark  \\  
Data Behaviour & $\times$  & \checkmark  & \checkmark \\  
Signature Detection &\checkmark & \checkmark  & \checkmark \\  
Device Action Perform & $\times$  & $\times$  & \checkmark \\ 
Detect Unknown Threats & $\times$  & \checkmark  & \checkmark \\
False Positive Ratio & Moderate  & High & Moderate \\ \hline
\end{tabular}
\end{center}
\label{Table-1: original with byzantine}
\end{table}
\end{comment}

\subsection{{Detection of faulty/ compromised Device(s)}}

In addition to determining a precise Oracle value, the proposed system offers the benefit of straightforward, self-sustaining accountability and auditability. In addition to pinpointing an accurate Oracle value, our proposed system simplifies the processes of accountability and auditability, making them inherently self-regulating. Within our framework, there's a built-in mechanism that keeps track of IoT devices with consistently errant data. These devices are flagged and their data is systematically excluded. Over a set time frame, devices that consistently submit incorrect readings are ranked. For each erroneous reading, a device-specific score (initially set to zero) increases. When a device's error rate reaches a threshold of $\varphi$, it's deemed malicious or compromised. Consequently, the system autonomously recommends its removal. This allows for the seamless replacement of malfunctioning devices with functioning ones. To implement this identification and replacement process, we've developed Algorithm 2. This algorithm is in the continuation of Algorithm 1 and uses the `Maximum attribute value' of IoT devices as an input criterion. The process unfolds as follows":
\begin{figure*}[ht]
\centerline{\includegraphics[height=7cm,width=0.7\textwidth]{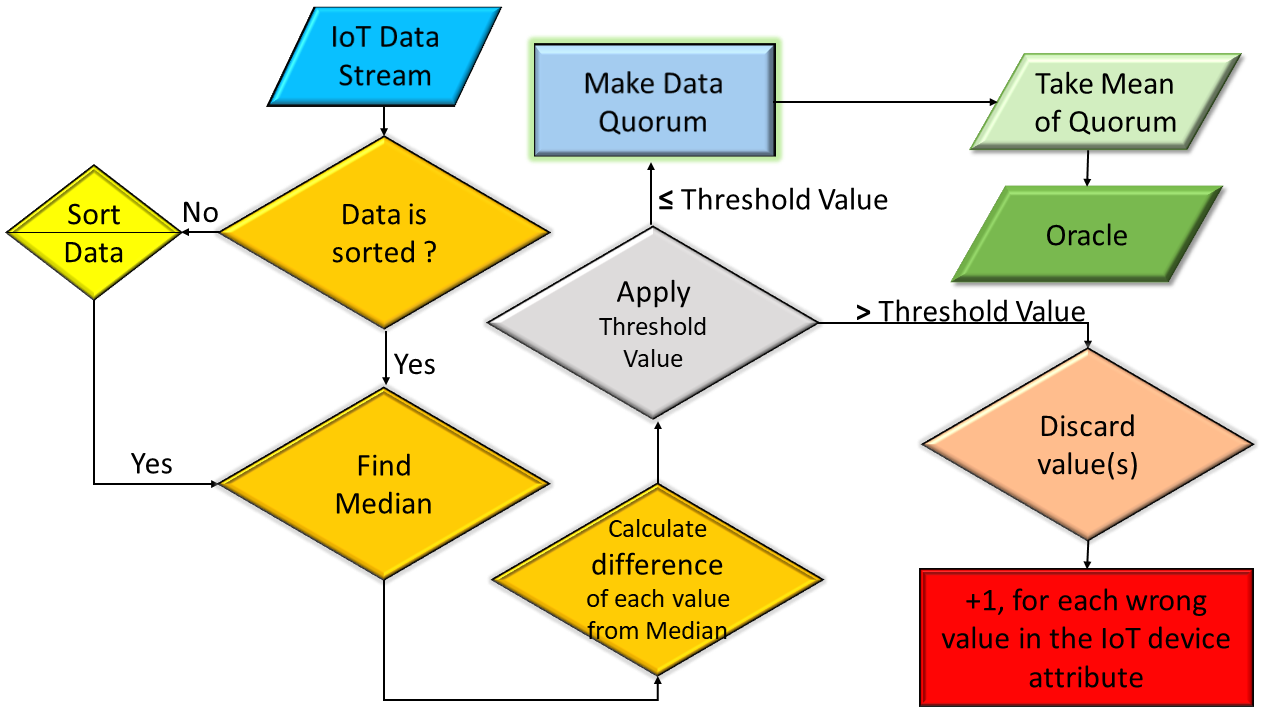}}
\caption{Information Flow Chart}
\label{fig2.1}
\end{figure*}

Before running this algorithm, it's essential to have an attribute value for every IoT device linked to the system. This attribute value, denoted as $(\varphi)$, represents the maximum count or score within a specified duration after which the data from that specific device will be discarded, spontaneously. At the outset, the attribute value of every device is initialized to $0$. The proposed system operates in the following manners:
\begin{enumerate}
\item Any values that are discarded are placed in a distinct array, referred to as 'Discarded values' ($Dsct$)
\item Initially, every device's attribute value starts at $0$
\item Whenever reading from an IoT device is discarded, the attribute value for that specific device increases by $1$
\item If an IoT device's attribute value hits the pre-set limit within the designated timeframe, the system will suspend accepting readings from that device
\item The system will then trigger an alert and flag this device as potentially compromised, suggesting it to be replaced
\end{enumerate}
% \vspace{0.1cm} 
  
\begin{table}[H]
\caption{Devices Score Board, where $\phi = 2$, $\Psi = 0$ and $\varphi = 5$}
\begin{tabularx}{0.47\textwidth}{ 
 | >{\centering\arraybackslash}X 
 | >{\centering\arraybackslash}X 
 | >{\centering\arraybackslash}X 
 | >{\centering\arraybackslash}X 
 | >{\centering\arraybackslash}X 
 | >{\centering\arraybackslash}X 
 | >{\centering\arraybackslash}X 
 | >{\centering\arraybackslash}X 
 | >{\centering\arraybackslash}X 
 | >{\centering\arraybackslash}X | }
 \hline
 & \cellcolor{gray} $d_1$ & \cellcolor{gray}  $d_2$ &  \cellcolor{gray} $d_3$ & \cellcolor{gray}  $d_4$ & \cellcolor{gray} $d_5$ & \cellcolor{gray} $d_6$ & \cellcolor{gray} $d_7$ & \cellcolor{gray} $d_8$ & \cellcolor{gray} $d_9$ \\
 \hline
i &   &   &   3 &     &    &   &   &  3 &  9 \\
 \hline
ii & - & - & $+1$ & - & - & - & - & $+1$ & $+1$ \\
 \hline
iii & - & - &\leavevmode\color{red}$\Uparrow$& - & - & - & - & \leavevmode\color{red} & \leavevmode\color{red}$\Uparrow$\\
 \hline
iv &  3  & -  & - &  -  &  -  & -  & -  & - &   9 \\
 \hline
v & $+1$ & - & - & - & - & - & - & - & $+1$ \\
 \hline
vi & \leavevmode\color{red}$\Uparrow$ & - & - & - & - & - & - & - & \leavevmode\color{red}$\Uparrow$ \\
\hline
vii &  3  &   &    &     &    &   &   &  &  11 \\
 \hline
viii & +1 & - & - & - & - & - & - & - & $+1$ \\
 \hline
ix & \leavevmode\color{red}$\Uparrow$ & - & - & - & - & - & - & - & \leavevmode\color{red}$\Uparrow$ \\
\hline
x & -  &   &    &     &    &   &   &  &  19 \\
 \hline
xi & - & - & - & - & - & - & - & - & $+1$ \\
 \hline
xii & - & - & - & - & - & - & - & - & \leavevmode\color{red}$\Uparrow$ \\
\hline
xiii & -  &   &    &     &    &   &   &  &  15 \\
 \hline
xiv & - & - & - & - & - & - & - & - & $+1$ \\
 \hline
xv & - & - & - & - & - & - & - & - & \leavevmode\color{red}$\Uparrow$ \\
\hline
 & \cellcolor{cyan}$\varphi=2$ & \cellcolor{green}$\varphi=0$ & \cellcolor{cyan}$\varphi=1$ & \cellcolor{green}$\varphi=0$ & \cellcolor{green}$\varphi=0$ & \cellcolor{green}$\varphi=0$ & \cellcolor{green}$\varphi=0$ & \cellcolor{cyan}$\varphi=1$ &\cellcolor{red}$\varphi=5$\\
\hline
\end{tabularx}
\label{Table-2: reliability}
\end{table}

In continuation of $Oracle\ calculation\ steps$, Table IV shows further steps ($i - xvi$) required to increase the attribute value of an $IoT$ device(s) and maintain $score$. Now in step $i$ we get discard and out of ranged values devices then assign $+1$ for discarded value device in each iteration in a specific time interval, subsequently in case of frequent wrong readings, device(s) will reach their score limit e.g. $\varphi=5$ at which the compromised device will be replaced.    

In Algorithm-2, consider $IoT_1, IoT_2, \ldots IoT_n$ as references to IoT devices connected to the system from remote locations. Each IoT device has an attribute value, ($Att_v$) denoted by $ \psi \in \mathbb{Z}$, where $\psi$ belongs to the set of integers $\mathbb{Z}$. Start by setting $ \psi $ to $0$ and define the attribute threshold as $\varphi$. During the process of examining the discarded values array ($Dtv$), for any discarded value originating from the $IoT_i$ IoT device, its attribute  $I_i$ gets incremented. Suppose the attribute value reaches $\varphi$ then readings from device $d\gets I$ corresponding to  $I_i$ are halted. Both algorithms can be visualized in Fig. 6.

\section{Simulations and Results}
\subsection{Data Set}
We utilized two Kaggle-published IoT temperature reading datasets\cite{kaggle2018} \cite{kagglecitydata2018}. These datasets aggregate temperature measurements from various IoT devices. The first dataset comprises approximately 77,000 temperature readings, gathered over four months from five distinct IoT devices. Conversely, the second dataset encompasses around 43,000 daily temperature readings, recorded over a span of more than four years from 10 different IoT devices. For our analysis, we handpicked a subset of 5,999 readings. Both datasets are accessible from GitHub repositories 
\footnote{https://github.com/Fahadrahman2121/IoT-temp-reading-}\&\footnote{https://github.com/Fahadrahman2121/temperature-city-data}.

\subsection{Evaluation}
For evaluation and comparison of ABQs model with different existing approaches, the following Equations are used i) Root Mean Square Error (RMSE), Equation 4, ii) Percent Error (PE), Equations 5, iii) Mean Absolute Error (MAE), Equations 6, iv) Mean Squared Error (MSE), Equations 7, v) Mean Absolute Percentage Error (MAPE), Equations 8, vi)  R-squared (Coefficient of Determination), Equations 9, vii) Adjusted $R^2$, Equation 10, viii) Mean Bias Deviation (MBD), Equations 11, ix) Median Absolute Deviation (MAD), Equations 12

{Root Mean Square Error $(RMSE)$}, Equations 4, 
\begin{align}
RMSE&= \sqrt{\frac{\sum(O_v - E_v)^2}{n}} 
\end{align} 
% where $O_v$ = observed-values, $E_v$ = expected-values. 

{Percent Error $(PE)$}, Equations 5
\begin{align}
PE  &= \frac{\sum(|O_v - E_v|)}{E_v} \times {100}
\end{align} 
where $O_v$ = Observed-values, $E_v$ = Expected-values.

{Mean Absolute Error $(MAE)$}, Equations 6
\begin{align}
MAE  &= \frac{1}{n} \sum_{i=1}^{n} |y_i - \hat{y_i}|
\end{align} 
where $y_i$ actual values, $\hat{y_i}$ predicted values, and $n$ is the number of observations.

{Mean Squared Error $(MSE)$}, Equations 7
\begin{align}
MSE  &= \frac{1}{n} \sum_{i=1}^{n} {({y_i - \bar{y_i})}^2}
\end{align}

{Mean Absolute Percentage Error  $(MAPE)$}, Equations 8
\begin{align}
MAPE &= \frac{100}{n} \sum_{i=1}^{n} |\frac{y_1 - \bar y_i}{y_1}|
\end{align}

{R-squared (Coefficient of Determination)}, Equations 9
\begin{align}
R^2  &= 1 - \frac{SS_{res}}{SS_{tot}}
\end{align} 
where $SS{_(res)}$ is the residual sum of squares $\sum{(y_i-\hat{y_i})}^2$ and $SS{_(tot)}$ is the total sum of squares $\sum {(y_i-\bar y_i)^2}$ with $\bar{y}$ as the mean of the observed data.

{Adjusted R-squared}, Equations 10
\begin{align}
Adjusted\ R^2 &= 1 - (1-R^2)\ \times\ \frac{n-1}{n-p-1}
\end{align} 
where $n$ represents the total data points in the dataset, and $p$ denotes the count of independent variables in the model.

{Mean Bias Deviation $(MBD)$}, Equations 11
\begin{align}
MBD  &= \frac{100}{n} \sum_{i=1}^{n} \frac{(y_i - \bar y_i)}{y_i}
\end{align}

{Median Absolute Deviation $(MAD)$}, Equations 12
\begin{align}
MAD  &= median (|y_1 - M|,|y_2-M|,...,|y_n - M|)
\end{align} 
where $M$ is the median of the dataset.
% \subsubsection{Logarithmic Loss (Log Loss)} Equations 16
% \begin{align}
% Log\ Loss  &= -\frac{1}{n} \sum_{i=1}^{n} [(y_i) * log( \bar y_i)] + (1-y_i) * log(1-\bar y_i)]
% \end{align} 
% where  $y_i$ is the true label (0 or 1), and $\bar y_i$ is the predicted probability of the instance being positive. 

Our simulation results highlight the efficiency of our proposed algorithm, showcasing its superior performance in ensuring data authenticity and the reliability of the Oracle value. Accurate readings closely align with the original data and contribute to the formation of ABQ. Furthermore, the accuracy of the ABQ system solutions improves with an increase in the number of IoT devices. It's also advisable to proportionally increase the $\phi$ value with the count of sensors. For comparisons of the data, evaluations, and visual representations, $Spyder\ (Python\ 3.9)$ and $Anaconda\ Navigator$ are used.

\subsection{Analysis of Statistical Hypotheses}
%\subsubsection{F-Test}
%F-Test is used to test the equality of two variances. It tells that the standard deviations from discussed  approaches are statistically different. The results of F-test are mentioned in Table I.   
%*************
%With reference to Table-I, Mean, Variance, F-value and P-values of existing and proposed approaches are calculated. According to F-test results, ABQ mean values results (for both data sets) are better than the other approaches. F-value is lesser than F-critical, and P-value is also less than 0.00. Hence in view of the results, ABQ approach is better in comparison with the other discussed approaches.  
%\begin{table}[t]
%\caption{F-Test Results For Data Set 1 \& 2}
%\begin{tabular}{m{0.8in}>
%{\centering\arraybackslash}m{0.4in}>
%{\centering\arraybackslash}m{0.7in}>
%{\centering\arraybackslash}m{0.3in}>
%{\centering\arraybackslash}m{0.4in}}
%\hline
%Technique & Mean & Variance &  F & P(F $\le$ f)  \\ 
%\hline
%& &{DATA SET 1} & &\\ 
%Mean & 41.088 & 6.361  & 0.849 & <0.00   \\ 
%Weighted & 41.688 & 6.467  & 0.863  & <0.00   \\  
%Consensus & 40.920 & 6.736  & 0.899 & <0.00   \\  
%ABQ & 40.321 & 7.486 & - & -  \\  \hline 
%& &{DATA SET 2} & &\\ 
%Mean & 81.286 & 140.774 & 0.849 & <0.00   \\  
%Weighted & 92.472  & 835.593 & 1.106  & <0.00  \\  
%Consensus & 81.133 & 146.296 & 1.038 & <0.00   \\  
%ABQ & 80.612 & 140.835 & -  & - \\  
%\hline
%\end{tabular}
%\label{Table-3: literature review}
%\end{table}
We conducted a t-test, which is based on t-distribution, to determine if there's a meaningful difference between the models. The resulting 
$p-values$ indicate a notable disparity (with 
($p$-value$<0.05$)  among the methodologies, as detailed in Table V.

\begin{table}[ht]
\caption{Comparing ABQ with Other Methods Using the t Test}
\begin{tabular}{m{1.7in}>
{\centering\arraybackslash}m{0.6in}>
{\centering\arraybackslash}m{0.6in}>
{\centering\arraybackslash}m{0.2in}}
\hline
Techniques & t value & $p$-value  \\ 
\hline
Mean approach         & 5.7 & < 0.0000   \\ 
Weighted approach     & 3.3 & < 0.0000  \\  
Consensus approach    & 3.1 & < 0.0000   \\  
\hline
\end{tabular}
\label{Table-3: T-Test Results}
\end{table}
\subsection{Reliability Analysis}
Constructing and operating economically crucial technological systems demand a thorough reliability assessment. Various methodologies systematically evaluate a system's reliability and risk. Fig. 7, shows Fault Tree Diagram (FTD), which is a widely adopted tool. Using the FTD derived from Equation 13, the reliability of the proposed Blockchain Oracle can be evaluated qualitatively through feedback on performance and dependability, and quantitatively by measuring specific metrics such as uptime, error rates, and response times.
\begin{align}
F(S) &= F(X_1)\ OR\ F(X_2) \ldots OR\ F(X_n) 
\end{align} 
where F($X_n$) is $n$-th event fail. 
% \begin{figure}[b]
% \centerline{\includegraphics[height=3cm,width=0.4\textwidth]{./Images/fai-.png}}
% \caption{Fault Tree Analysis of proposed model}
% \label{fig7}
% \end{figure}
\begin{figure}[b]
\centerline{\includegraphics[height=3cm,width=0.4\textwidth]{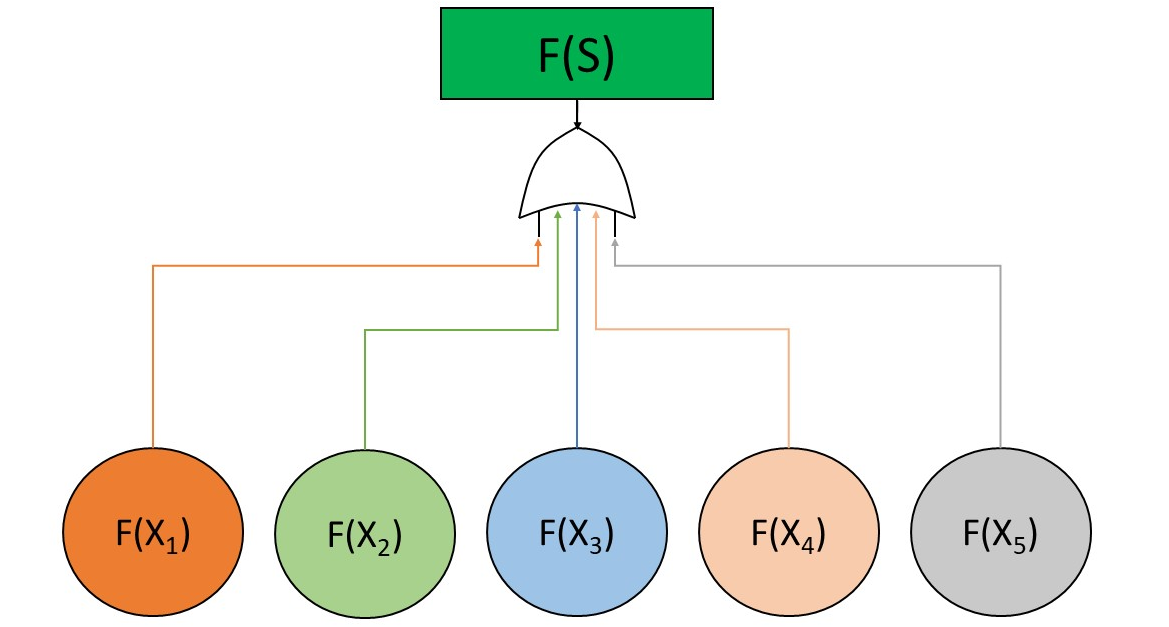}}
\caption{Fault Tree Analysis of proposed model}
\label{fig7}
\end{figure}

In this system, we have experienced that reliability $(r)$ is enhanced when $sensors$ are connected in 'parallel' (${par}$), instead of connections in series. This means that the system's reliability improves with an increase in the number of sensor(s). The reliability of our proposed system, denoted as ($R_{par}$) is determined by Equation 14.

\begin{align}
R_{par} & = 1 - (1 - r)^m 
\end{align} 
Where, $r$ represents the reliability of a single unit, while $m$ denotes the count of active units. Given that an IoT device has a reliability of 0.966, a minimum of five IoT devices is required to achieve 100\% system reliability. The detailed reliability metrics for the system are presented in Table VI.

\begin{table}[H]
\caption{System Reliability in the Proposed Model}
\begin{center}
\begin{tabular}{m{1.0in}>
{\centering\arraybackslash}m{1.3in}>
{\centering\arraybackslash}m{0.2in}}
\hline
Number of Devices & Reliability-Analysis   \\ 
\hline
1 & 0.9660000        \\ 
2 & 0.9984000      \\  
3 & 0.9999360    \\  
4 & 0.9999974   \\  
5 & 0.9999999  \\   
\hline
\end{tabular}
\end{center}
\label{Table-2: reliability}
\end{table}
\subsection{Absolute Improvement}
This is simply the difference between the accuracy of the new technique and the accuracy of the old technique, as shown in Equation 15. Table VII presents a comparative analysis with ABQ, revealing absolute improvement of 2.6, 1.5, and 1.4 with the Weighted, Consensus, and Mean methods, respectively
\begin{align}
Absolute\ Improvement &= A_{new}- A_{old} 
\end{align} 
where $A_{new}$ is accuracy of ABQ  technique and $A_{old}$ is accuracy of Weighted, Consensus and Mean techniques.
\begin{table}[ht]
\caption{Absolute Improvement}
\begin{center}
\begin{tabular}{m{1.0in}>
{\centering\arraybackslash}m{0.6in}>
{\centering\arraybackslash}m{0.2in}}
\hline
Technique & Absolute Improvement   \\ 
\hline
ABQ Vs Weighted  & 2.6   \\ 
ABQ Vs Consensus & 1.5   \\  
ABQ Vs Mean      & 1.4   \\  
\hline
\end{tabular}
\end{center}
\label{Table-3: Absolute Improvement}
\end{table}
\subsection{$F_1$ Score}
The $F1$ score serves as an indicator of a model's precision and recall balance, providing a composite accuracy assessment in machine learning tasks. Its value spans between $0$ (least optimal, the worst) and $1$ (most optimal) with higher values indicating better accuracy and completeness in predictions. For the calculation of $F_1$ score, first, we need to find precision and recall values. 
\subsubsection{Precision and Recall Evaluation} 
Precision assesses the correctness of positive predictions, whereas Recall gauges the model's capability to detect all pertinent cases. Both metrics range between 0 and 1, with higher values indicating superior performance. Equations 16 and Equation 17 are used to find the values of Precision and Recall, respectively.  The results are shown in Table VIII. Where True Positive is abbreviated as $TP$ and False Positive as $FP$.
\begin{align}
Precision  &= \frac{TP}{TP\ +\ FP} \\
Recall  &= \frac{TP}{TP\ +\ FP} 
\end{align} 
% , and $FN$ is false negatives.
\begin{table}[ht]
\caption{Precision and Recall Values}
\begin{center}
\begin{tabular}{m{1.0in}>
{\centering\arraybackslash}m{0.8in}>
{\centering\arraybackslash}m{0.8in}}
\hline
Prediction Method & Precision & Recall   \\ 
\hline
ABQ-Value          & 1.0000 & 0.9849 \\ 
Mean-Value         & 1.0000 & 0.9843 \\  
Consensus-Value    & 0.9988 & 0.9896  \\ 
Weighted-Value     & 0.9993 & 0.9705  \\ 
\hline
\end{tabular}
\end{center}
\label{Table-3: Precision and Recall Values}
\end{table}
\subsubsection{$F_1$ Score Evaluation} 
For the evaluation of $F_1$ score, Equation 18 is executed on dataset 1 and dataset 2, the results are mentioned below in Table IX:
\begin{align}
F_1  &= 2\ \times\ \frac{precision\ \times\ recall}{precision\ +\ recall}  
\end{align} 
% where precision and recall are defined next.
\begin{table}[ht]
\caption{F1 Score}
\begin{center}
\begin{tabular}{m{0.8in}>
{\centering\arraybackslash}m{0.4in}>
{\centering\arraybackslash}m{0.4in}>
{\centering\arraybackslash}m{0.4in}>
{\centering\arraybackslash}m{0.4in}>
{\centering\arraybackslash}m{0.1in}>
{\centering\arraybackslash}m{0.0in}}
\hline
&  \multicolumn{2}{c}{Dataset 1} & \multicolumn{2}{c}{Dataset 2}\\
\midrule
Prediction Method & $F_1$ Score & Accuracy & $F_1$ Score & Accuracy  \\  
\hline 
ABQ-Value       & 0.9964 & 0.9890 & 0.9994 & 0.9895 \\ 
Mean-Value      & 0.9921 & 0.9844 & 0.9921 & 0.9844 \\  
Consensus-Value & 0.9922 & 0.9855 & 0.9922 & 0.9885 \\  
Weighted-Value  & 0.9847 & 0.9700 & 0.9847 & 0.9700 \\ 
\hline
\end{tabular}
\end{center}
\label{Table-3: Absolute Improvement}
\end{table}
The outcome of Table IX is based on threshold values $30$ and $60$ for $dataset_1$ and $dataset_2$, respectively:
\begin{itemize}
\item $ABQ-Value$ has the highest $F1$ Score and accuracy at these threshold values.
\item $Mean-Value$ and $Consensus-Value$ follow closely behind.
\item $Weighted-Value$ has a slightly lower $F1$ Score and accuracy in comparison.
\end{itemize}
Dataset 1 and dataset 2 are applied to the above equations and inferred the results in graphical format which are depicted in Fig. 8 to Fig. 16. The results are clearly showing that the accuracy and performance of the proposed $ABQ$ approach is far better than the existing approaches.

The discussed statistical techniques i.e. $RMSE$, $PE$, $(MSE)$, $(MAE)$, $R^2$, $(MBD)$, $(MAPE)$, Adjusted $R^2$, and $(MAD)$ are mathematical measures used to analyze and interpret data, providing insights and supporting decision-making based on empirical evidence. From descriptive statistics to advanced inferential predictive modeling, it is proved that the accuracy of the $ABQ$ approach is much better than the compared related work. This paper concludes with the key factors of a Blockchain Oracle in comparison with related published work in Table X.

% \begin{figure}[H]
% \centerline{\includegraphics[height=5cm,width=0.5\textwidth]{./Images/RMSE1-.png}}
% \caption{Root Mean Square Error of different approaches for data set 1 and 2 are shown in (a) and (b), respectively}
% \label{fig13.1}
% \end{figure}
\begin{figure}[H]
\centerline{\includegraphics[height=5.5cm,width=0.5\textwidth]{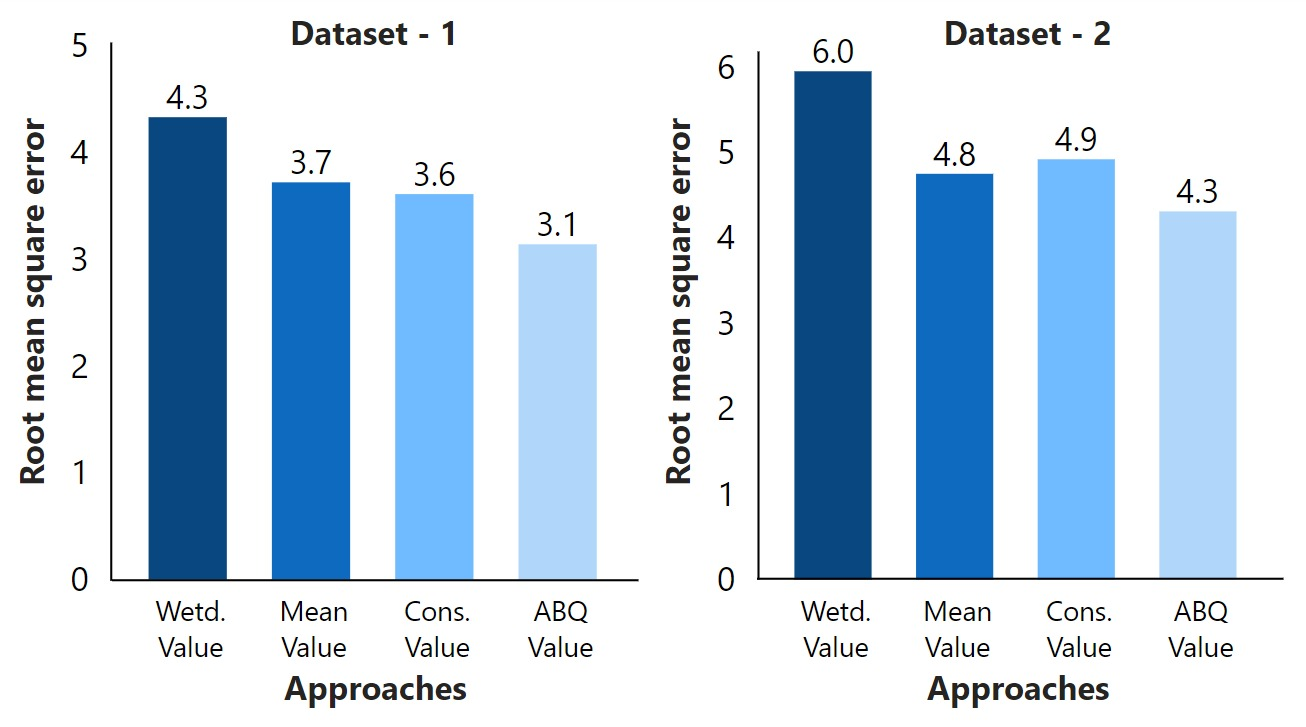}}
\caption{Results of RMSE for dataset-1 and dataset-2}
\label{fig13.1}
\end{figure}

\vspace{1cm}

\begin{figure}[H]
\centerline{\includegraphics[height=5.5cm,width=0.5\textwidth]{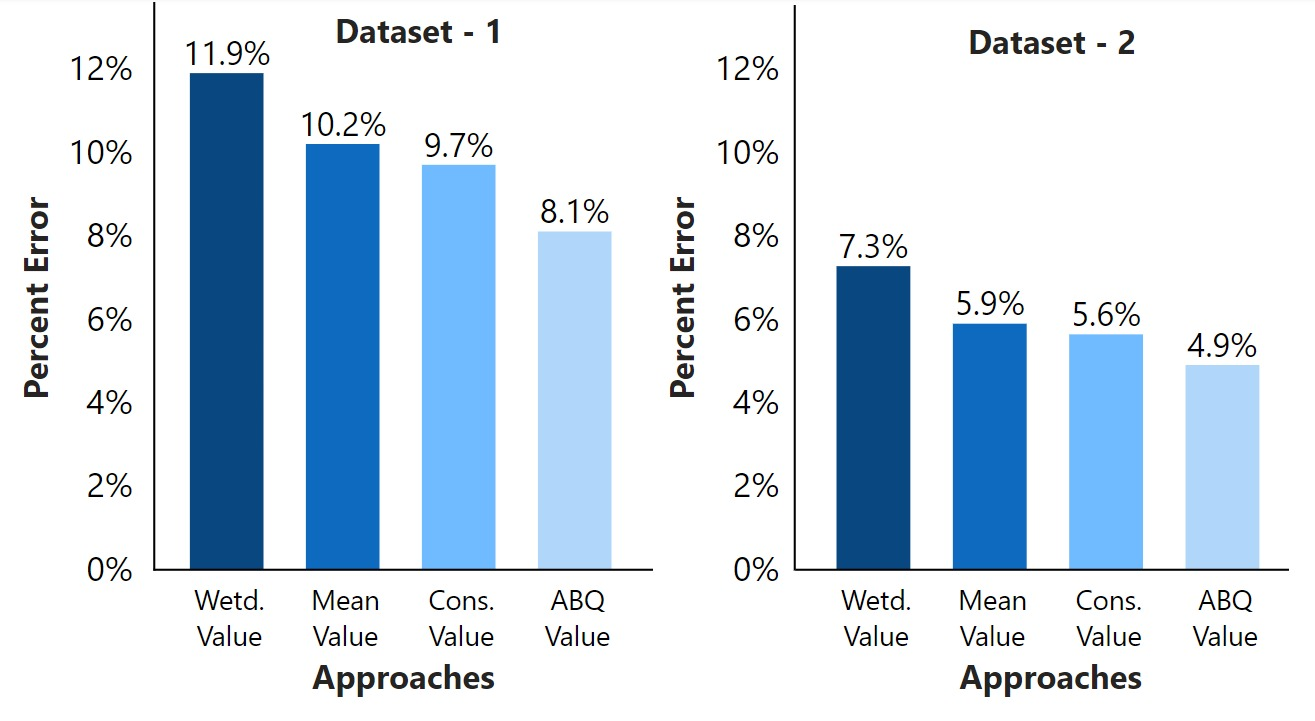}}
\caption{Results of PE for dataset-1 and dataset-2}
\label{fig10.1}
\end{figure}

% \vspace{1cm}

\begin{figure}[H]
\centerline{\includegraphics[height=5.5cm,width=0.5\textwidth]{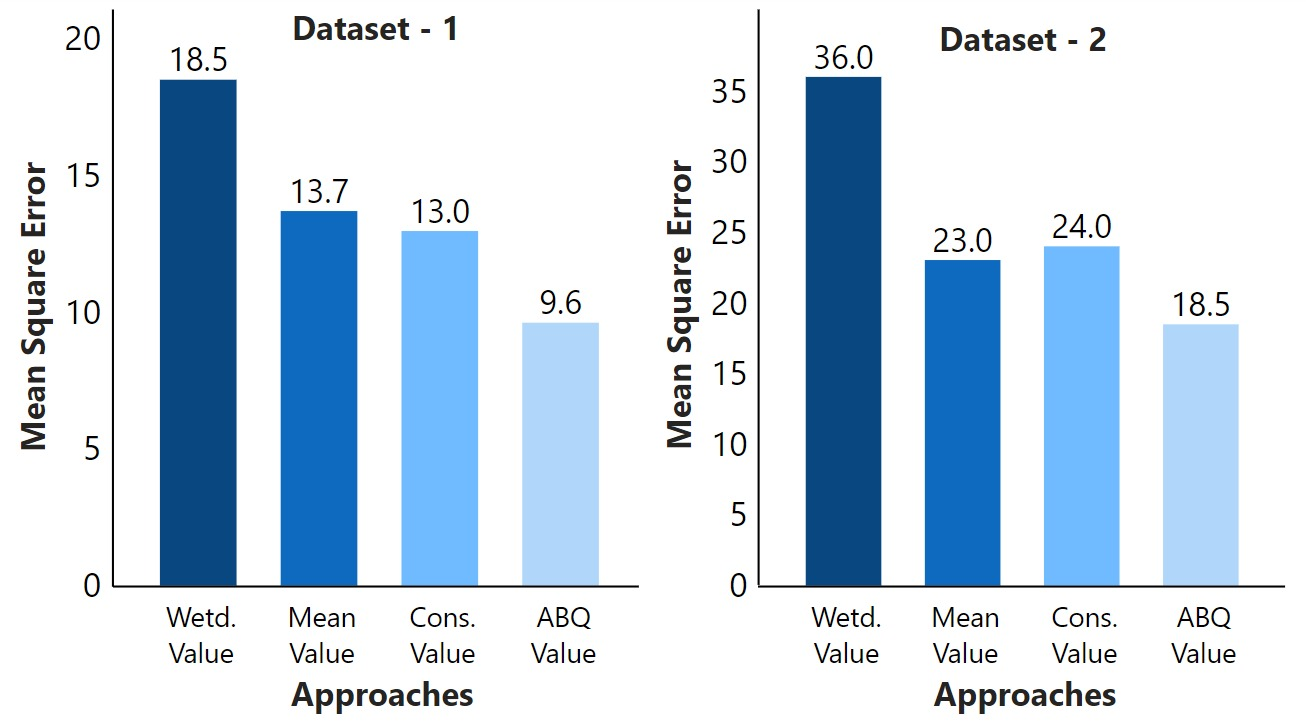}}
\caption{Results of MSE for dataset 1 and dataset 2}
\label{fig13.1}
\end{figure}

\begin{figure}[H]
\centerline{\includegraphics[height=5cm,width=0.5\textwidth]{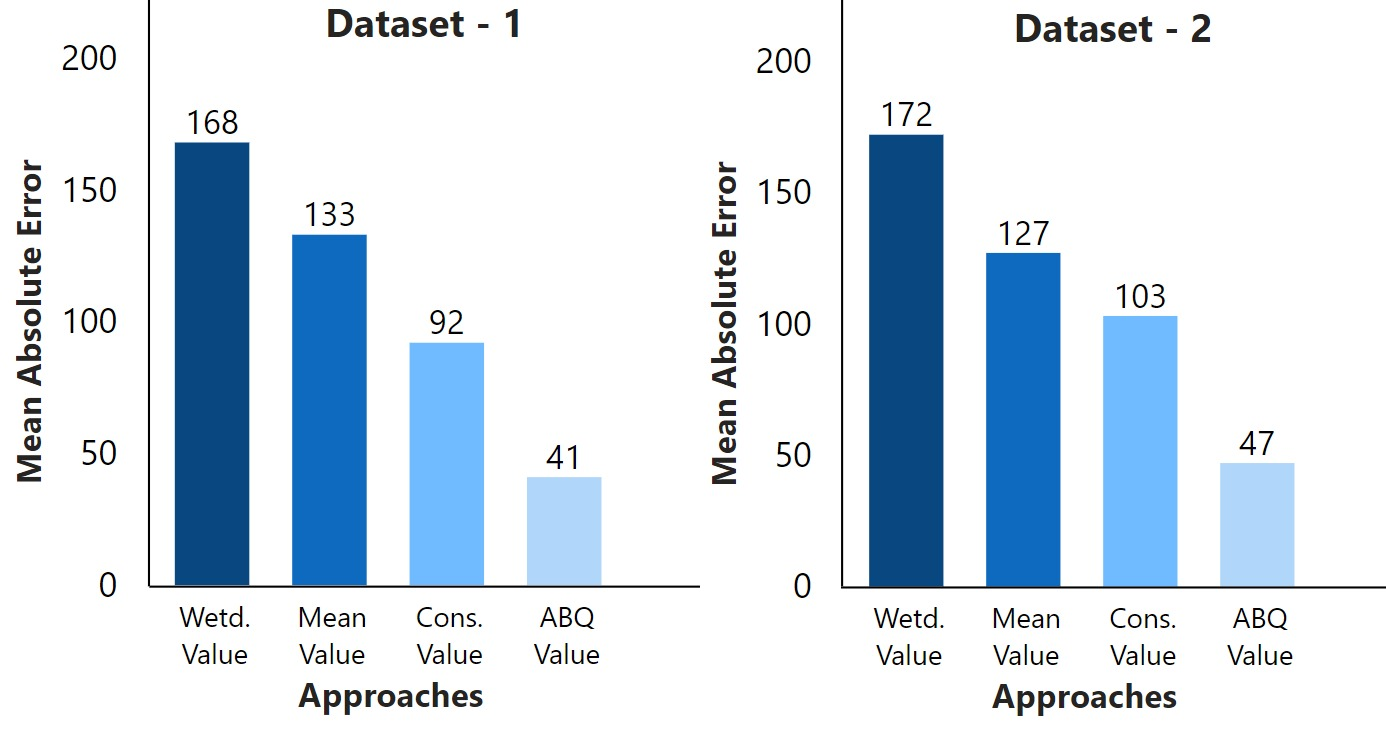}}
\caption{Results of MAE for dataset-1 and dataset-2}
\label{fig13.1}
\end{figure}

\begin{center}

\begin{figure}[H]
\centerline{\includegraphics[height=5cm,width=0.5\textwidth]{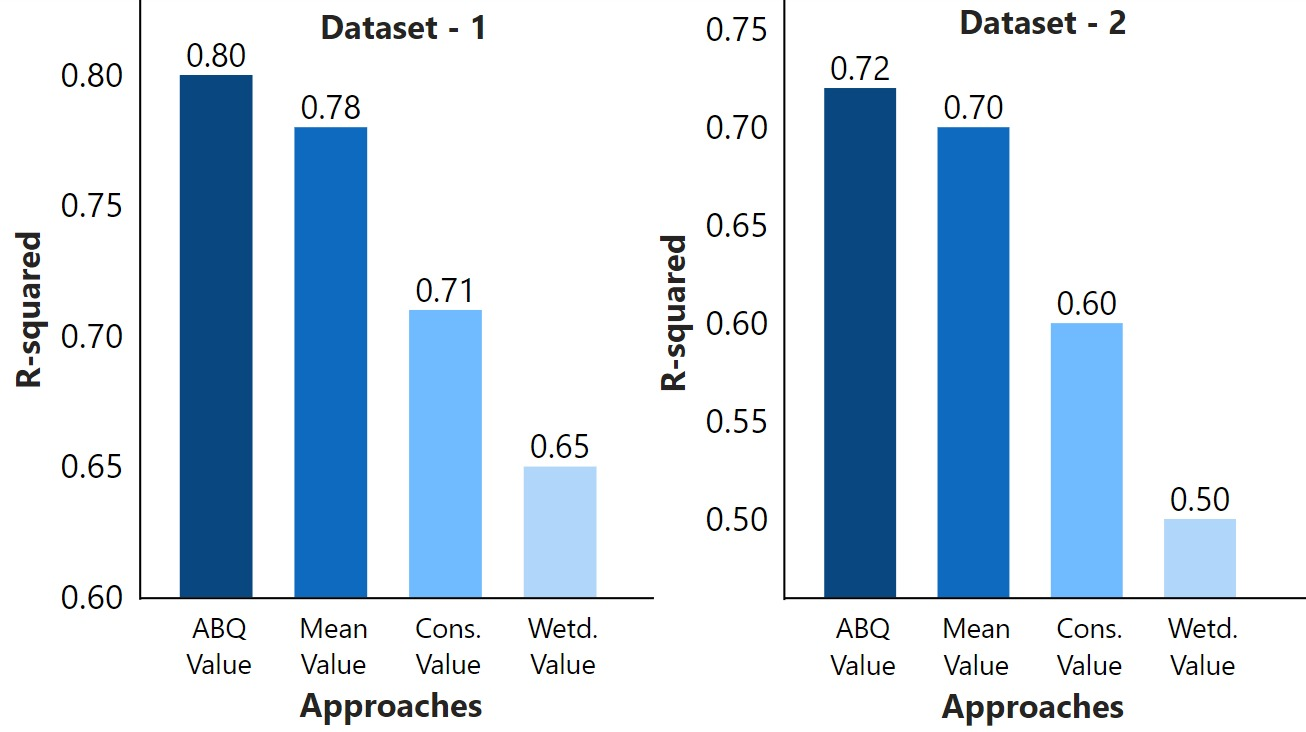}}
\caption{Results of R-Squared for dataset-1 and dataset-2}
\label{fig13.1}
\end{figure}

\end{center}

\begin{figure}[b]
\centerline{\includegraphics[height=5cm,width=0.5\textwidth]{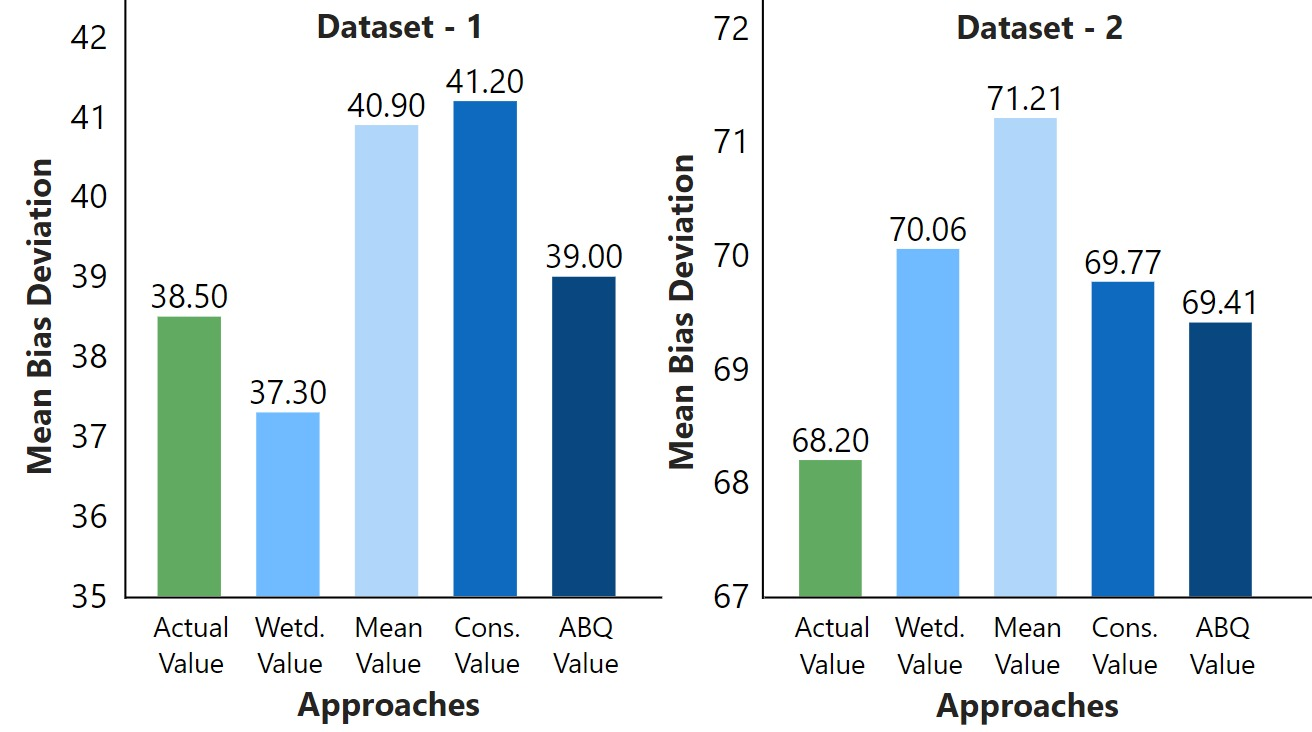}}
\caption{Results of MBD for dataset-1 and dataset-2}
\label{fig13.1}
\end{figure}

\begin{figure}[H]
\centerline{\includegraphics[height=5cm,width=0.5\textwidth]{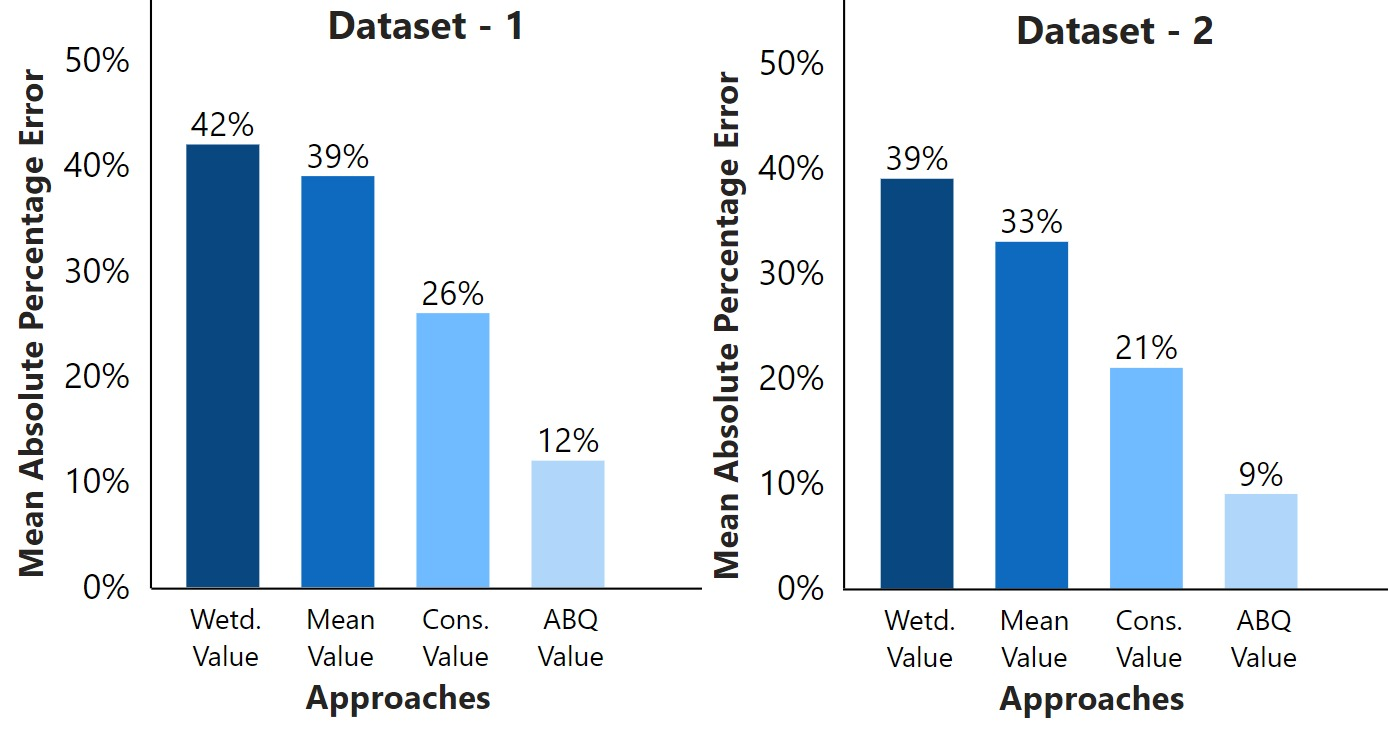}}
\caption{Results of MAPE for dataset-1 and dataset-2}
\label{fig13.1}
\end{figure}

\vspace{1.7cm}

\begin{center}

\begin{figure}[H]
\centerline{\includegraphics[height=5cm,width=0.5\textwidth]{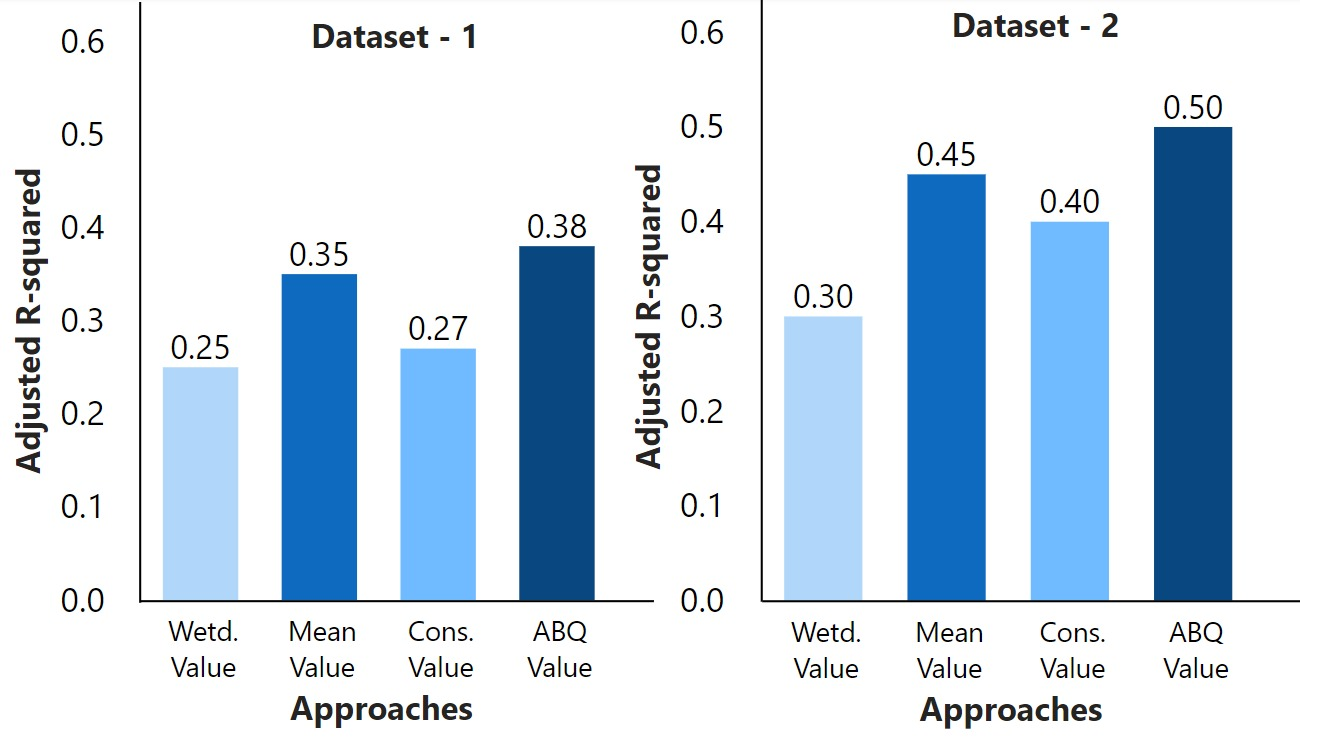}}
\caption{Results of Adjusted R-Squared for dataset-1 and dataset-2}
\label{fig13.1}
\end{figure}

\end{center}

\begin{figure}[b]
\centerline{\includegraphics[height=5cm,width=0.5\textwidth]{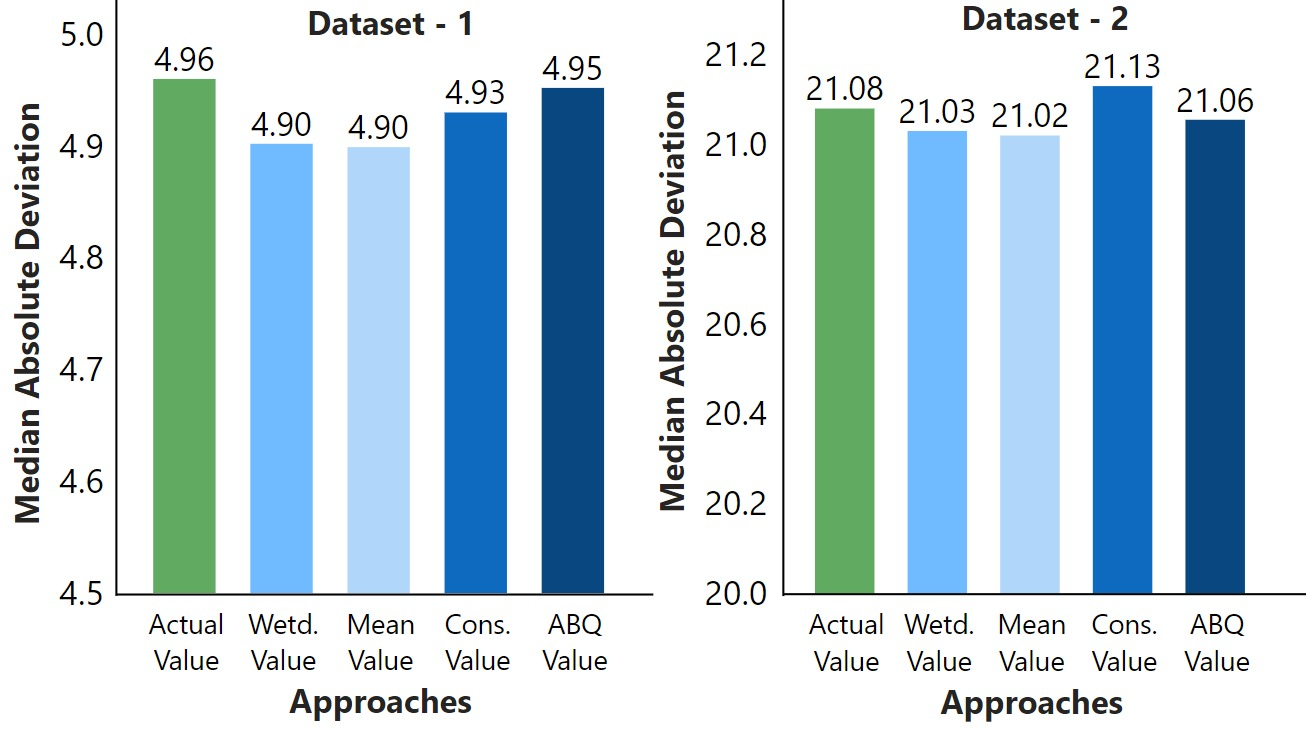}}
\caption{Results of MAD for dataset-1 and dataset-2}
\label{fig13.1}
\end{figure}
 
\begin{table*}[t]
\caption{Proposed model in Comparison with related work}
\begin{tabular} {|p{3cm}|p{2cm}|p{2cm}|p{2cm}|p{2cm}|p{2cm}|p{2cm}|}
\hline
Authors & Security & Simplicity & Trustlessness &  Transparency & Redundancy & Independent\\ 
\hline
S. Ellis et al.~\cite{ellis2017chainlink} & No & No & No & No & No & No \\ 
Adler et al.~\cite{adler2018astraea} & Yes & No & No & No & No & No   \\ 
Tian et al.~\cite{tian2019evaluating} & No & No & No & No & No & No   \\ 
% Gautam et al.~\cite{srivastava2019light}            & Yes & Yes  & No & No & Medium  \\ 
Heiss et al.~\cite{heiss2019oracles} & Yes & No & No & No & No & No   \\ 
Berger et al.~\cite{berger2020aware} & Yes & No & No & No & No  & No  \\ 
Tseng et al.~\cite{tseng2020exact} & No & No & No & No & No & No  \\ 
Wei et al.~\cite{wei2020pspl} & Yes & No & No &  No & No & No \\ 
Proposed Solution  & Yes & Yes & Yes & Yes & Yes & Yes  \\ 
\hline
\end{tabular}
\label{Table-3: literature review}
\end{table*}

\vspace{2cm}

\section{Conclusion and Future Work}

In this study, a novel approach is introduced in which ABQs and heuristic-based detection are clubbed together for accuracy and pinpointing compromised devices, respectively. Our method leans on the foundational Blockchain belief that the majority (two-thirds) of nodes are consistently accurate. Through our methodology, we observed the generation of highly accurate and trustworthy values, facilitating the swift identification of any malfunctioning data-transmission units. We deduced that $ABQs$ offer a more streamlined and rapid solution for extracting near-authentic values from dubious data sources. The Heuristic-based detection ($HBD$) stands out as an efficient tool for spotting faulty nodes, demonstrating resilience by maintaining operations even when over a quarter of its nodes fail or exhibit malicious behavior. Our empirical findings reveal that the precision of the $ABQs$ method surpasses and offers greater resilience than traditional Blockchain Oracle techniques. Malicious or compromised nodes can be promptly pinpointed through the audit and accountability mechanisms embedded in the `heuristic-based' detection approach.

In the future, Machine Learning (ML) can be employed to design quicker and more efficient Oracles. Considering data trails or backtracking data to its Oracle value could improve the autonomy of Blockchain Oracle.

\bibliographystyle{unsrt}
\bibliography{ref.bib}

\vspace{12pt}
\color{red}
% IEEE conference templates contain guidance text for composing and formatting conference papers. Please ensure that all template text is removed from your conference paper prior to submission to the conference. Failure to remove the template text from your paper may result in your paper not being published.

\end{document}